\documentclass[aps,prb, superscriptaddress, linenumbers%
aip,
amsmath,amssymb,
reprint,%
]{revtex4-2}
\usepackage[
pdffitwindow=true,
colorlinks=true,
frenchlinks=false,
linkcolor=blue,
anchorcolor=blue,
citecolor=blue,
filecolor=blue,
urlcolor=blue,
bookmarks=true,
bookmarksopen=true,
bookmarksnumbered=true,
bookmarksopenlevel=1,
plainpages=false,
pdfpagelayout=TwoPageLeft,
pdfpagelabels=true,
breaklinks
]{hyperref}
\usepackage[per-mode=symbol,separate-uncertainty]{siunitx}
\usepackage{graphicx}
\usepackage{dcolumn}
\usepackage{booktabs}
\usepackage{url}
\usepackage{color}
\usepackage{graphicx,wrapfig,lipsum}
\usepackage{bm}
\usepackage[utf8]{inputenc}
\usepackage[english]{babel}
\usepackage{color}
\usepackage{ulem}
\usepackage{amssymb}
\usepackage{dsfont}
\usepackage{lipsum}
\bibpunct{[}{]}{,}{n}{}{}
\usepackage{braket}
\usepackage{chemformula}
\usepackage{soul}
\usepackage[dvipsnames]{xcolor}

\usepackage{tabularx}
\newcolumntype{Y}{>{\centering\arraybackslash}X}

\begin{document}
	\title{Multi-cavity strong coupling to an electron spin ensemble: \\ spectral and dark-state signatures}
	\author{P.~Oehrl}
    \altaffiliation{These authors contributed equally.}
	\affiliation{Walther-Mei{\ss}ner-Institut, Bayerische Akademie der Wissenschaften, Garching, Germany}
	\affiliation{TUM School of Natural Sciences, Technical University of Munich, Garching, Germany}
    \author{B.~P\'erez Gonz\'alez}
    \altaffiliation{These authors contributed equally.}
	\affiliation{Institute of Physics, University of Augsburg, Augsburg, Germany}
	\author{A.~Dunaev}
	\affiliation{Walther-Mei{\ss}ner-Institut, Bayerische Akademie der Wissenschaften, Garching, Germany}
	\affiliation{TUM School of Natural Sciences, Technical University of Munich, Garching,  Germany}
	\author{M.~Althammer}
	\affiliation{Walther-Mei{\ss}ner-Institut, Bayerische Akademie der Wissenschaften, Garching, Germany}
	\affiliation{TUM School of Natural Sciences, Technical University of Munich, Garching, Germany}
    \author{T.~S.~Parvini}
	\affiliation{Walther-Mei{\ss}ner-Institut, Bayerische Akademie der Wissenschaften, Garching, Germany}
    \author{F.~Piazza}
    \affiliation{Institute of Physics, University of Augsburg, Augsburg, Germany}
    \author{M.~Benito}
    \affiliation{Institute of Physics, University of Augsburg, Augsburg, Germany}
    \affiliation{Center for Advanced Analytics and Predictive Sciences, University of Augsburg, Augsburg, Germany}
	\author{H.~Huebl}
	\email{huebl@wmi.badw.de}
	\affiliation{Walther-Mei{\ss}ner-Institut, Bayerische Akademie der Wissenschaften, Garching, Germany}
	\affiliation{TUM School of Natural Sciences, Technical University of Munich, Garching, Germany}
	\affiliation{Munich Center for Quantum Science and Technology (MCQST), Munich, Germany}
	\begin{abstract}
    Spin ensembles are considered as potential candidates for quantum memory and quantum enhanced sensing applications. Here, we explore the controlled coupling of multiple superconducting microwave cavities to a spin ensemble, which shows signatures of strong coupling and, due to the multi-mode character, the formation of dark states. In particular, the latter are of interest, as they provide a potential pathway to enhance memory times and enable protected storage of non-classical states in spin ensembles due to the suppressed coupling to the circuit environment. We model the spin multi-cavity hybrid to reproduce the spectra and extract characteristic coupling strengths using the input-output formalism.  
	\end{abstract}
	\maketitle

\section{Introduction}

The generation of non-classical states in magnetic excitations, such as squeezed and entangled states, is central to quantum-enhanced sensing and to the realization of so-called quantum tokens for quantum communication protocols\,\cite{li2022perspective, huang2024entangl, zhou2025entangl, Hillery2000, gavinsky2012quantum, Fesquet2024, oehrl2025}. In this context, paramagnetic spin ensembles are particularly attractive because of their long coherence times \cite{morton2008nature, morley2010nature, steger2012science, saeedi2013science, dantec2021science, strinic2025prb} and their demonstrated compatibility with superconducting microwave circuits\,\cite{bushev2011prb, probst2013prl}. In cavity-based implementations, coherent coupling to an electromagnetic mode provides both a means to manipulate and read out the spin ensemble and an interface to other quantum degrees of freedom\,\cite{PhysRevLett.102.083602, PhysRevLett.110.250503, PhysRevLett.107.060502, oehrl2025, Xiang2013}. Such platforms also provide access to multi-mode spin dynamics and to interference phenomena associated with the collective nature of the ensemble\,\cite{putz2017naturephoton, putz2017circuit, Grezes2014}.\\
Going beyond a single cavity mode considerably extends these capabilities. Multi-mode hybrid systems can support multiple competing excitation pathways, and provide additional control parameters for state transfer, entanglement generation, and reservoir engineering\,\cite{groszkowski2022prx, wooley2014pra, andersen2012pra, liu2016scirep, mao2021apl}. Particularly appealing is the possibility of multi-partite strong coupling\,\cite{mao2025genuinetripartitestrongcoupling, PhysRevResearch.3.023093, lee2026}, which enables coherent excitations to be distributed among distinct subsystems with complementary functionalities, as light-matter coupling is distributed among collective modes.\\
An additional consequence of multi-mode hybridization is the emergence of dark states. More generally, dark states are collective eigenmodes whose coupling to a particular degree of freedom is suppressed by coherent interference. Depending on the physical mechanism, this can render them inaccessible to a given external probe or relaxation channel, thereby suppressing radiative losses and potentially enhancing their coherence properties\,\cite{Okamoto:2016bm, doi:10.1073/pnas.1419326112, an2022bright, Zanner2022}. The term \emph{dark state} encompasses several physically distinct situations. In $\Lambda$-type systems, coherent population trapping can suppress occupation of a common excited state through destructive interference \cite{xu2008coherent}. The same dark-state interference underlies electromagnetically induced transparency and dark-state polaritons \cite{harris1997electromagnetically, fleischhauer2002pra, Lukin2003colloq}, with analogous phenomena in optomechanical systems \,\cite{harris1997electromagnetically, weis2010science, zhou2013nature}. Dark collective modes can also arise through hybridization-induced decoupling
from an intermediate subsystem or external probe
\,\cite{putz2017naturephoton, li2020hybrid}, particularly at multipartite resonances
\,\cite{an2022bright, zhang2015magnon, mao2025genuinetripartitestrongcoupling}.\\
Of particular relevance to circuit and waveguide quantum electrodynamics are radiatively dark states associated with a common electromagnetic environment. In these systems, several modes can couple to the same transmission line and therefore share a radiative decay channel. The different components of a collective excitation then radiate coherently into the common continuum, allowing their emission amplitudes to interfere destructively\,\cite{Zanner2022, Propp22}. When this cancellation is complete, the collective state becomes dark with respect to the corresponding environmental channel.\\
Here, we investigate both multi-mode strong coupling and radiative dark-state formation in a hybrid microwave system composed of a diphenyl-picrylhydrazyl (DPPH) spin ensemble and three superconducting resonators coupled to a common transmission line. Two of the resonators are close in frequency and coherently hybridize, whereas the spin ensemble couples directly to only one of them. The resonator-resonator interaction nevertheless redistributes the spin-photon coupling over both hybridized photonic modes. We show that both branches satisfy their respective strong-coupling conditions, demonstrating a genuine tripartite spin-resonator-resonator strong-coupling regime in which the resonator detuning and inter-resonator coupling control the effective spin-photon coupling, linewidth, and cooperativity of each branch.\\
At the same time, the common transmission line provides a shared radiative environment for the resonators, allowing the fields emitted by the different components of a hybridized eigenmode to interfere in the output channels. Using an input-output description of the device, we derive the conditions under which an eigenstate becomes dark to transmission, reflection, or to both channels simultaneously. Crucially, the experimentally observed disappearance of the central resonance from the transmission spectrum provides a direct signature of this interference mechanism: for the fitted system parameters, the corresponding hybridized eigenstate satisfies the condition for vanishing radiation into the transmitted channel.\\
These results show that the same multi-mode hybridization responsible for tripartite strong coupling also provides a means to engineer the coupling of collective excitations to their electromagnetic environment. In contrast to dark states generated by avoiding population of an intermediate mode or by imposing a multipartite resonance condition, the darkness considered here originates directly from destructive interference between radiative amplitudes emitted into a shared continuum. This provides a route towards mode-selective readout and radiative protection in spin-resonator architectures based on linear multi-mode coupling and magnetic-field control of the hybrid-mode composition.

\begin{figure}[!t]
    \centering
    \includegraphics[width=1\linewidth]{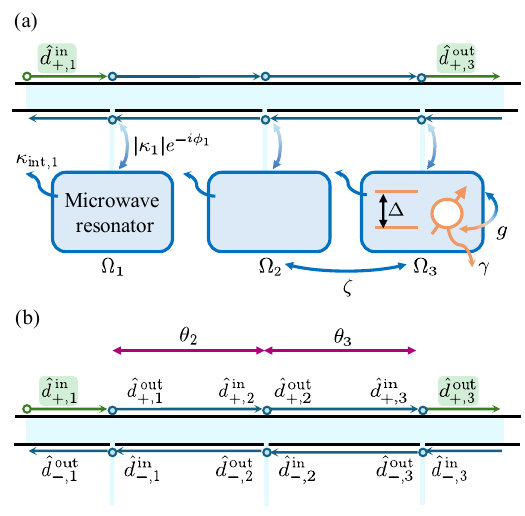}
    \caption{\textbf{Conceptual depiction of the setup.} \textbf{(a)} The superconducting microwave circuit consists of three hanger-type, lumped-element resonators of frequencies $\Omega_1$, $\Omega_2$, $\Omega_3$, which are coupled to a common transmission line with coupling strengths $\vert \kappa_i \vert e^{-i\phi_i}$. One of the resonators is strongly coupled, with coupling strength $g$, to the spin ensemble with resonance frequency $\Delta$. We investigate the complex transmission parameter $S_{21}=\braket{\hat{d}^{\text{out}}_{+, 3}}/\braket{\hat{d}^{\text{in}}_{+, 1}}$ in order to study multi-mode interference and dark-state formation. We also consider the dissipation rates $\kappa_{\text{int},i}$ and $\gamma$ for the resonators and the spin ensemble, respectively. \textbf{(b)} Left and right propagation channel through the transmission line.} 
    \label{fig1:setup}
\end{figure}

\section{Model Hamiltonian \label{sec:hamiltonian}}
In our experiment, we investigate a superconducting circuit composed of three planar, hanger-type resonators that are capacitively coupled to a common transmission-line waveguide, as shown in  Fig.\,\ref{fig1:setup}(a), modeled by the following Hamiltonian \cite{gardiner1985input,steck2007quantum,chiorescu2010magnetic}
\begin{align}
\label{eq: Hamiltonian}
\begin{split}
    \hat{\mathcal{H}} = \hat{\mathcal{H}}_{\text{cav-cav}} + \hat{\mathcal{H}}_{\text{spin}} +\hat{\mathcal{H}}_{\text{spin-cav}}  + \hat{\mathcal{H}}_{\text{cav-tl}}\text{.}
\end{split}
\end{align}
The first term in the Hamiltonian describes the microwave resonators and their mutual interaction,
\begin{align}
\label{Eq: cav cav interaction}
\begin{split}
    \hat{\mathcal{H}}_{\text{cav-cav}}/\hbar &= \sum_{j = 1}^3 \Omega_j \hat{a}^\dagger_j \hat{a}_j + \sum_{i,j}\zeta_{ij} (\hat{a}_i \hat{a}^{\dagger}_j + \hat{a}^{\dagger}_i \hat{a}_j ) \text{,}
\end{split}
\end{align}
where $\Omega_j$ corresponds to the resonator frequencies, with $\hat{a}_j^{(\dagger)}$ being their corresponding bosonic annihilation (creation) operators. We assume a finite coupling strength $\zeta_{ij}$ between the $i^{\text{th}}$ and $j^{\text{th}}$ cavity in the circuit, which can originate from direct electromagnetic cross-talk\,\cite{yan2023apl, chen2023npj}, or from a broad dissipative bath mode associated with the device packaging, which provides an effective channel that mediates the interaction. Due to the frequency detunings in our system, $\Omega_1 \ll(\Omega_2,\Omega_3)$ and $\Omega_2 \sim \Omega_3$, the couplings to the first resonator are largely suppressed, $\zeta_{12},\, \zeta_{13} \approx 0$\,\cite{PhysRevA.93.042307}, so that we only consider a finite coupling $\zeta_{23} \equiv \zeta$ in the following.

The spin ensemble  is provided by paramagnetic moments with  $S=1/2$ hosted in DPPH molecules. Their collective dynamics are well described within the single-excitation manifold as\,\cite{wesenberg2009quantum}:
\begin{align}
\begin{split}
    \hat{\mathcal{H}}_{\text{spin}}/\hbar &= \frac{\Delta}{2}\hat{S}_z\text{,}
\end{split}
\end{align}
where $\hat{S}_z = \sum_{i=1}^N \hat{\sigma}_{z,i}$ is the collective spin operator, and $\hat{\sigma}_{z,i}$ describe identical spin operators for each of the $N$ spins of the ensemble. Notice, that we are neglecting the sub-ensemble distribution of spin-resonance frequencies in our model, which is a good approximation in the single-excitation limit\,\cite{PhysRev.114.1219, chiorescu2010magnetic, PhysRevA.84.063810, PhysRevX.7.041011}. The ensemble is exclusively coupled to the third resonator (\textcolor{black}{see Appendix \ref{app: Appendix Experimental Setup} for details}), described in the form of the Tavis-Cummings interaction Hamiltonian\,\cite{taviscummings1968pr}
\begin{align}
\begin{split}
    \hat{\mathcal{H}}_{\text{spin-cav}}/\hbar &= g (\hat{a}_3 \hat{\sigma}_+ + \hat{a}^{\dagger}_3 \hat{\sigma}_- ) \text{,}
\end{split}
\label{eq:spincavint}
\end{align}
where $g=\sqrt{N}g_0$ is the collective coupling strength obtained from the single-spin coupling rate $g_0$ and the number of spins $N$.

Finally, the last term in the Hamiltonian describes the transmission line mode, through which we can externally drive the circuit\,\cite{probst2015efficient,PhysRevB.106.214506,rieger2023fano}
\begin{align}
\begin{split}
\label{eq: Hamiltonian cav-tl}
    \hat{\mathcal{H}}_{\text{cav-tl}}/\hbar &=  \sum_{j=1}^{3} \sum_{p=\pm}\int_0^{\infty} \text{d}\nu \nu \hat{d}^{\dagger}_{p,j}(\nu) \hat{d}_{p,j}(\nu) \\
 &- i \sum_{j=1}^{3} \sum_{p=\pm}\int_0^{\infty} \text{d}\nu \big[ \sqrt{\frac{\kappa_j}{2\pi}} \hat{a}^{\dagger}_j \hat{d}_{p,j}(\nu) -  \text{h.c.} \big] \text{.}
\end{split}
\end{align}
The coupling between the $j^{\text{th}}$ intra-cavity mode and the external microwave circuit is described via the complex coupling rate $\kappa_j=|\kappa_j|\text{e}^{-i\phi_j}$, where the phase accounts for imperfections of the microwave circuit, such as impedance mismatches, reflections between the input and output ports of our waveguide, or a finite path length between the spatial locations of the microwave cavities\,\cite{khalil2012analysis,deng2013analysis,probst2015efficient}. The annihilation and creation operators for the microwave modes are defined as $\hat{d}_{p,j}$ and $\hat{d}^{\dagger}_{p,j}$, respectively, and satisfy $[\hat{d}_{p,j}(\nu), \hat{d}^\dagger_{p^\prime,j}(\nu^\prime)] = \delta_{p,p^\prime} \delta_{j,j^\prime} \delta(\nu - \nu^\prime)$. The two subindices account for the propagation direction $p = \pm$ (for right- and left -moving fields, respectively) within the waveguide, and the cavity $j$ that they are coupled to. 
While the coherent dynamics of the light-matter system is governed by the first 
three 
terms in the Hamiltonian, Eqs.\,\eqref{Eq: cav cav interaction}-\eqref{eq:spincavint}, the explicit inclusion of the transmission line Hamiltonian of Eq.\,\eqref{eq: Hamiltonian cav-tl} is important to properly model the emergence and detection of a stationary dark state in this multi-mode setup.

\begin{figure*}[!tb]
    \centering
    \includegraphics[width=1\textwidth]{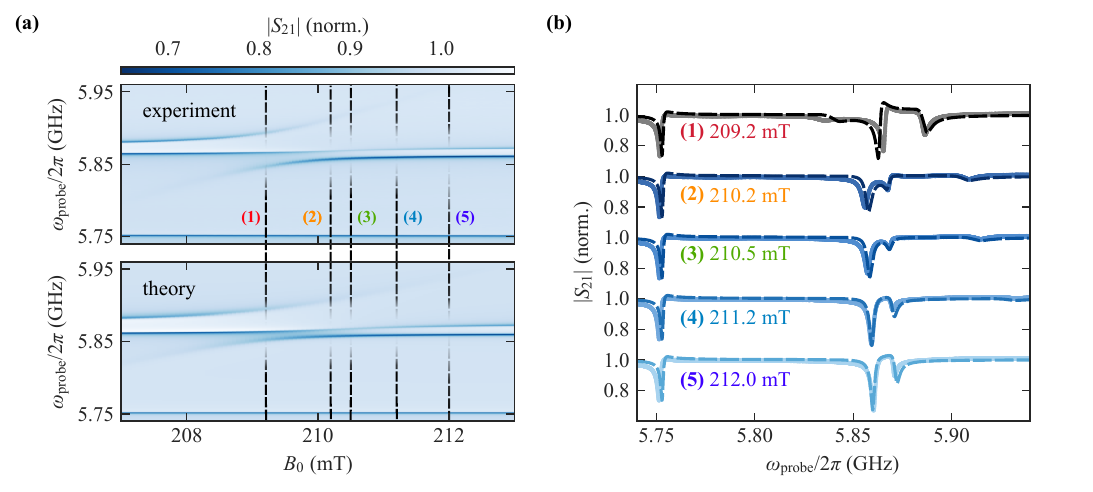}
    \caption{\textbf{Measured and modeled cwESR spectra.} \textbf{(a)} Measured (upper panel) and simulated (lower panel) microwave absorption amplitude $|S_{21}|$ as a function of the applied magnetic field $B_0$ and the probe frequency $\omega_{\text{probe}}$. The modeled microwave absorption amplitude is based on Eq.\,\ref{eq: Hamiltonian} and the coupling rates summarized in Tab.\,\ref{tab:spin_hamiltonian}. Vertical lines $(1)-(5)$ represent $B_0-$fixed cuts represented in panel (b). \textbf{(b)} Comparison of the measured (solid line, light colored) and modeled (dashed line, dark colored) data showing the frequency-dependent microwave absorption amplitude for fixed magnetic fields.}
    \label{fig2}
\end{figure*}

\section{Results}
We perform continuous wave electron spin resonance (cwESR) experiments at a temperature of $\SI{4}{\kelvin}$ by measuring the complex microwave transmission $S_{21}$ as function of the probe frequency $\omega_\mathrm{probe}$ and the applied static magnetic field $B_0$ (\textcolor{black}{see Appendix \ref{app: Appendix Experimental Setup} for details}). 
Figure\,\ref{fig2}\,(a) presents the measured cwESR spectrum recorded at an input microwave power of $P_{\text{mw}}=-55$~dBm~$\approx\SI{3}{\nano\watt}$. We corrected the data to account for the complex microwave background and offsets in the static magnetic field (\textcolor{black}{see Appendix \,\ref{Appendix: background correction}}). 
The spectrum reveals three distinct, field-independent horizontal absorption modes, which originate from the fixed-frequency microwave cavity resonances of the resonators 1, 2, and 3 with frequencies $\Omega_1/(2\pi)=\SI{5.753}{\giga\hertz}$, $\Omega_2/(2\pi)=\SI{5.864}{\giga\hertz}$, and $\Omega_3/(2\pi)=\SI{5.873}{\giga\hertz}$, respectively. 
We determine the resonance frequencies by fitting the steady-state solution of the theoretical input-output model to the measured spectrum at the lowest magnetic field, assuming no interaction with the spin ensemble at this point and a finite cavity-cavity coupling, which is determined to be $\zeta_{23}/(2\pi)=\SI{6.0}{\mega\hertz}$.  

Moreover, we observe a field dependent cwESR absorption signature, which is characteristic for a paramagnetic spin ensemble. This mode exhibits a linear relation between the resonance frequency and the applied magnetic field and displays a pronounced anti-crossing behavior as shown in Fig.\,\ref{fig2}. Notably, the spin mode hybridizes with both resonator 2 and 3, despite the spin ensemble being physically located  only on resonator 3. Additionally, at about $\SI{210}{\milli\tesla}$ and $\SI{5.87}{\giga\hertz}$, we experimentally observe a pronounced suppression of the absorption signature, which is indicative of the formation of a dark state.

To characterize the system quantitatively, we fit the measured response using the input–output model\,\cite{gardiner1985input, steck2007quantum, xiao2010asymmetric, PhysRevB.106.214506}, which we derive from the multi-element Hamiltonian in Eq.\,\eqref{eq: Hamiltonian}. 
We obtain the classical input-output relations in the steady-state regime by defining the input fields \(\hat d^{\mathrm{in}}_{p,j}\) and output fields \(\hat d^{\mathrm{out}}_{p,j}\) associated with cavity \(j\) for both propagation directions \(p=\pm\) (\textcolor{black}{see Appendix\,\ref{Appendix: details of theory}}). We refer the reader to Fig.~\ref{fig1:setup}(b) for a schematic. To account for the propagation delay of the microwave modes between the resonator coupling points, we introduce an additional phase \(\theta_j=\nu \Delta x_j / v_{\mathrm g}\), where \(\nu\) is the angular frequency, \(v_{\mathrm g}\) is the group velocity, and \(\Delta x_j\) is the propagation distance between resonators $\Delta x_j = \vert x_j - x_{j-1}\vert $\,\cite{xiao2010asymmetric,PhysRevB.106.214506}. 
This leads to the following relations between the input and output microwave fields:
\begin{align}
\begin{split}
    \hat{d}^\mathrm{in}_{-,1} &= e^{i\theta_2}\hat{d}^\mathrm{out}_{-,2}\text{,}
    \qquad
    \hat{d}^\mathrm{in}_{+,2} = e^{i\theta_2}\hat{d}^\mathrm{out}_{+,1}\text{,}\\
    \hat{d}^\mathrm{in}_{-,2} &= e^{i\theta_3}\hat{d}^\mathrm{out}_{-,3}\text{,}
    \qquad
    \hat{d}^\mathrm{in}_{+,3} = e^{i\theta_3}\hat{d}^\mathrm{out}_{+,2}\text{.}
\end{split}
\label{eq:input_output_withphase}
\end{align}
We also include the spin dephasing rate \(\gamma\) and the cavity loss rate \(\kappa_{\mathrm{cav},j}\), which accounts not only for the dissipation induced by the transmission line, but also for additional decay channels that may cause further broadening of the cavity linewidth. Using these relations, we obtain the complex transmission coefficient $S_{21}=\braket{\hat{d}^{\mathrm{out}}_{+,3}}/\braket{\hat{d}^{\mathrm{in}}_{+,1}}$, which we use to extract the parameters of the multi-element system. 

\begin{table*}[!tb]
    \centering
    \caption{\textbf{Summary of the fitted coupling parameters.}
    To extract the characteristic parameters of our hybrid system, we fit the experimentally recorded spectrum with the input–output formalism based on Eq.\,\ref{eq: Hamiltonian}.}
    \label{tab:spin_hamiltonian}
    \begin{tabular}{*{9}{c}}
    \toprule
    $\Omega_1/(2\pi)$ & $\Omega_2/(2\pi)$ & $\Omega_3/(2\pi)$ 
    & $\kappa_{\mathrm{cav},1}/(2\pi)$ & $\kappa_{\mathrm{cav},2}/(2\pi)$ & $\kappa_{\mathrm{cav},3}/(2\pi)$  
    & $\kappa_1/(2\pi)$ & $\kappa_2/(2\pi)$  & $\kappa_3/(2\pi)$ \\
    \midrule
    5.753\,GHz & 5.864\,GHz & 5.873\,GHz 
    & 1.488\,MHz & 1.650\,MHz & 2.872\,MHz 
    & 0.114\,MHz & 0.211\,MHz & 0.297\,MHz \\
    \midrule
    $\phi_1$ & $\phi_2$ & $\phi_3$  
    & $\theta_2$ & $\theta_3$ 
    & $g/(2\pi)$ & $\gamma/(2\pi)$ & $\zeta_{23}/(2\pi)$ & \\
    \midrule
    -0.50\,rad & -0.92\,rad & 0.95\,rad 
    & 0.4\,rad & 1.8\,rad 
    & 20.0\,MHz & 12.0\,MHz & 6.0\,MHz & \\
    \bottomrule
    \end{tabular}
\end{table*}


Using the calibrated cavity parameters, we extract the collective spin--cavity coupling and the spin relaxation rate from a two-dimensional fit to the hybridized region ($B_0=205$ mT to $215$ mT). The fit is performed by minimizing the corresponding cost function with the L-BFGS-B algorithm, using bounded parameter ranges. The fitted parameters of the multi-element system are summarized in Tab.~\ref{tab:spin_hamiltonian}. As a result, the lower panel of Fig.\,\ref{fig2}\,(a) shows the modeled spectrum, which agrees closely with the measured data. To provide further detail, we present fixed-field line cuts that compare the measured and modeled absorption amplitude as a function of the probe frequency. The modeled data accurately describes the experimental results, as shown in Fig.\,\ref{fig2}\,(b).

\section{Discussion}
Within this section, we discuss two specific signatures of the measured spectrum in more detail: the strong coupling regime of the multi-mode system, and the formation of the dark state. \\

\begin{figure}[!t]
    \centering
    \includegraphics{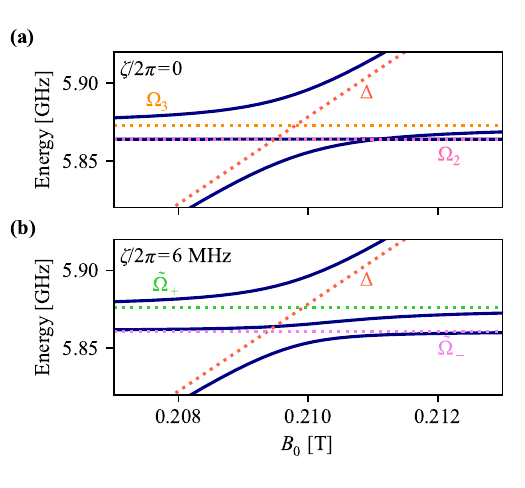}
    \caption{\textbf{Simulated energy spectrum for the tripartite system.}
    Calculated energies of the effective three-component system formed by the spin ensemble and resonators 2 and 3, as a function of $B_0$, for the fitted parameters in Table \ref{tab:spin_hamiltonian}, comparing \textbf{(a)} $\zeta/2\pi = 0$, \textbf{(b)} $\zeta/2\pi = 6$ MHz. The dotted lines indicate the bare energies of the modes that are hybridizing: \textbf{(a)} \(\Delta\), \(\Omega_2\), and \(\Omega_3\), \textbf{(b)} \(\Delta\), $\tilde{\Omega}_+$, and $\tilde{\Omega}_-$.}
    \label{fig:lambdasystem_energyspectrum}
\end{figure}

\textbf{Multi-mode strong coupling.} 
The measured and simulated transmission spectra indicate that the device operates in a multi-mode strong-coupling regime. The observed splittings cannot be interpreted as an isolated avoided crossing between a bare cavity mode and the collective spin excitation, instead, they originate from the hybridization of the spin ensemble with the normal modes of the coupled-resonator subsystem. Within the few excitation limit \cite{wesenberg2009quantum, chiorescu2010magnetic}, the direct inter-cavity coupling $(\zeta)$ hybridizing resonators 2 and 3 can be described by
\begin{eqnarray}
    \ket{\tilde{\psi}_+} & = & \cos(\varphi/2)\ket{1,0} + \sin(\varphi/2)\ket{0,1}, \label{eq:eigenstates_plus}\\
    \ket{\tilde{\psi}_-} & = & -\sin(\varphi/2)\ket{1,0} + \cos(\varphi/2)\ket{0,1}\label{eq:eigenstates_minus},
\end{eqnarray}
where $\ket{n_2, n_3}$ indicates the number of excitations in resonator 2 or 3, respectively, and $\tan(\varphi) = 2\zeta/(\Omega_2 - \Omega_3)$, with the corresponding energies
\begin{equation}
    \tilde{\Omega}_{\pm} = \frac{\Omega_+}{2}\pm \frac{1}{2}\sqrt{\Omega_-^2 + 4\zeta^2},
    \label{eq:twores_eig}
\end{equation}
using $\Omega_\pm = \Omega_2 \pm \Omega_3$. 
As the spin ensemble couples directly to resonator 3, both hybrid photonic modes acquire a finite coupling to the spin ensemble. 
In the basis $\{\ket{\downarrow, \tilde{\psi}_+},\ket{\downarrow, \tilde{\psi}_-}, \ket{\uparrow,0,0}\}$, the cavity-cavity-spin Hamiltonian takes the form
\begin{equation}
    \hat{\mathcal{H}}_{\text{s-r-r}} = \left( 
        \begin{array}{ccc}
              \tilde{\Omega}_+ - \Delta & 0 & g\sin\left(\frac{\varphi}{2}\right) \\
            0 &  \tilde{\Omega}_- -  \Delta  &  g\cos\left(\frac{\varphi}{2}\right) \\
            g\sin\left(\frac{\varphi}{2}\right)  &        g\cos\left(\frac{\varphi}{2}\right)  & 0
        \end{array}
    \right),
    \label{eq:hamiltonian_spin_hybrized}
\end{equation}
where $\ket{\tilde{\psi}_-}$ and $\ket{\tilde{\psi}_+}$ are now coupled to the spin, with coupling strengths $g_{+} = g\sin(\varphi/2)$ and $g_{-} = g\cos(\varphi/2)$, respectively. The energy spectrum of the tripartite system is illustrated in Fig.~\ref{fig:lambdasystem_energyspectrum}. Notice that there are two distinct Rabi splittings, for both resonance conditions $\tilde{\Omega}_+ = \Delta$ and $\tilde{\Omega}_- = \Delta$. \\

In a two-component system consisting of a photonic mode and a matter excitation, the onset of strong coupling is typically identified by the condition \(g \gtrsim (\kappa+\gamma)/2\), meaning that the light-matter coupling strength must overcome the combined losses of the photonic and matter subsystems. For \(\zeta=0\), resonator 2 acts as a passive spectator to the light-matter hybridization, and the conditions for the onset of the strong-coupling regime can be expressed entirely in terms of the parameters of resonator 3 and the spin ensemble. 
 
For $\zeta \neq 0$, however, each of these couplings $g_-$ and $g_+$ give rise to a distinct Rabi splitting in the energy spectrum, and one may therefore formulate a separate strong-coupling condition for each branch. To do so, it is essential to determine the effective decay rates of the hybridized modes \(\ket{\tilde{\psi}_\pm}\) in terms of the bare decay rates of resonators 2 and 3. One way to address this issue is to directly solve the non-Hermitian Hamiltonian accounting for both coherent dynamics and dissipation, and analyze its eigenvalues. Following this method, we find that the corresponding effective linewidth for $\ket{\tilde{\psi}_\pm}$ is (\textcolor{black}{see Appendix \ref{sec:app_decaylength_twores}}) ,
\begin{align}
    \tilde{\kappa}_{+/-} & = \kappa_{\mathrm{cav},2/3}\cos^2\left(\frac{\varphi}{2}\right) + \kappa_{\mathrm{cav},3/2}\sin^2\left(\frac{\varphi}{2}\right) \nonumber \\
    &  \hspace{15pt} \pm  2\sqrt{\vert \kappa_2 \kappa_3 \vert }\cos\left(\theta_3 - \frac{\phi_2 + \phi_3}{2} \right)\sin\left( \varphi \right).
    \label{eq:linewidth_resonantmodes}
\end{align}
In this estimation, we neglected the effect of resonator 1, as it is far detuned from the pair $(2,3)$ and the corrections provided are of second order in the cavity linewidths, $\propto \kappa_1\kappa_j / (\Omega_1 - \Omega_j)$. Notice that, apart from the expected contribution of the original decay rates of resonators 2 and 3 $\kappa_{\mathrm{cav}}$, i.e., first line of Eq. \eqref{eq:linewidth_resonantmodes}, the phases $\theta_3$, $\phi_2$, and $\phi_3$, also modulate the linewidth of the hybridized modes, as shown in the second line. For the parameters shown in Table \ref{tab:spin_hamiltonian}, we can estimate a decay rate of $\tilde{\kappa}_+ / 2\pi \approx 2.54$ MHz, and $\tilde{\kappa}_- /2\pi \approx 1.98$ MHz for each mode, together with the corresponding light-matter coupling strengths of $g_+ / 2\pi \approx 17.9$ MHz and $g_- / 2\pi \approx 8.9$ MHz. With this, we can confirm not only the strong coupling regime of the spin ensemble to resonator 3, $g > (\kappa_{\mathrm{cav},3} + \gamma)/2$, but also to both hybridized photonic modes with $g_{\pm} > (\tilde{\kappa}_\pm + \gamma)/2$. With this, a cooperativity of $C_\pm = 4g_{\pm}^2/(\tilde{\kappa}_{\pm}\gamma)$ can be estimated for each Rabi splitting, yielding $C_+ \approx 41.95$ and $C_- \approx 13.47$. Therefore, in tripartite systems composed of two resonators and one spin system, the detuning $\Omega_-$ and the inter-resonator coupling $\zeta$ act as control parameters that redistribute the spin oscillator strength between the two hybridized photonic branches. This can be used to selectively enhance the cooperativity of one branch while suppressing unwanted hybridization with the other, providing a route to mode-selective strong coupling, optimized readout, or the generation of protected dark-mode configurations, as we will show below.\\

\textbf{Dark state generation.} The measurements also show a suppressed absorption dip as a function of $B_0$ for the central branch, which indicates the formation of a stationary dark state. In our hybrid system, the tripartite strong coupling creates multiple coherent channels for excitation exchange. In case of destructive interference, the resulting hybridized mode becomes effectively decoupled from the input–output ports of the microwave circuit, which appears experimentally as the disappearance of the absorption dip. Certain hybridized states thus become effectively shielded from dissipation and external probing, consistent with the observed suppressed transmission feature. This behavior confirms that the system supports coherent superpositions that selectively decouple from the measurement ports, providing direct evidence for dark‑state generation within the multi‑mode hybridized manifold.

In general, the generation of dark states based on multi-mode interference requires a radiation of the involved modes into a common continuum\,\cite{Propp22}. For each propagation direction, $p = \pm$, the output field can be decomposed into a freely propagated input contribution and a system-dependent radiative source operator $\hat{L}_p$, such that 
\begin{equation}
    \hat{d}^{\mathrm{out}}_{p,i} = e^{i\Theta}\hat{d}^{\mathrm{in}}_{p,j} + \hat{L}_p,\quad p = \pm,
\end{equation}
where the input and output ports $j,i$ depend on the channel $p$, as shown in Fig. \ref{fig1:setup}(b), and $\Theta$ is the total propagation phase accumulated over the full length of the circuit. A stationary eigenstate of the coherent Hamiltonian $\ket{\Psi_\alpha}$ is dark, when its radiated field vanishes in the relevant output channel, $\hat L_p\ket{\Psi_\alpha}=0$ \cite{Stannigel_2012, RevModPhys.97.025004, PhysRevA.91.042116, PhysRevA.78.042307, PhysRevB.102.115109}. If this condition is satisfied for both $p = \pm$, the state is decoupled from the common feedline and is therefore dark in both transmission and reflection. \\
%
%

\begin{figure}
    \centering
    \includegraphics{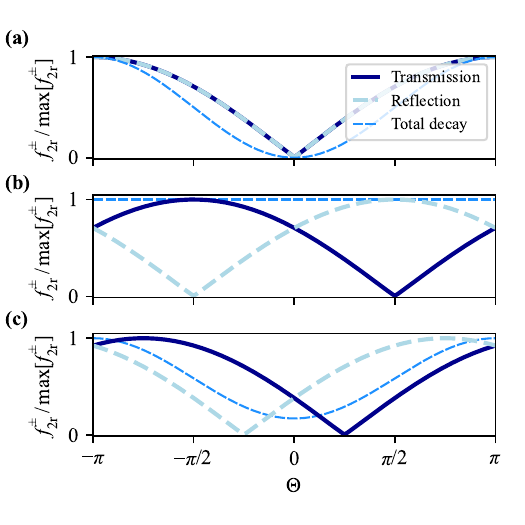}
    \caption{\textbf{Dark and directionally dark modes in the two-coupled-resonator system.} Dark state condition $f^\pm_{\text{2r}} / \mathrm{max}[f^\pm_{\text{2r}}]$ for $\ket{\tilde{\psi}_+}$, as a function of the propagation phase $\Theta$, for complex $\kappa_i = \vert \kappa_i\vert e^{-i\phi_i}$, and for both propagation channels $\hat{L}_\pm$. The diagnostic function $F_{\text{2r}}/\mathrm{max}[F_{\text{2r}}]$ is included. A bidirectional dark state appears for $f^{\pm}_{\text{2r}} = 0$. The system parameters $\Omega_{2,3}$, $\zeta$, and $\kappa_{2,3}$, are chosen such that they satisfy Eq. \eqref{eq:magnitude_twores}. \textbf{(a)} The phases $\phi_2 = 3\pi/2$ and $\phi_3 = -\pi/2$ fulfill $\vert \phi_3 - \phi_2\vert = 2\pi$ ($l = 1$), so a dark state to both $p = \pm$ appears at $\Theta = 0$. \textbf{(b)} The phases $\phi_2 = 3\pi/2$ and $\phi_3 =\pi/2$, and \textbf{(c)} $\phi_2 = 7\pi/4$ and $\phi_3 = \pi/4$, are chosen such that the dark state condition happens at a different $\Theta$ for transmission and reflection. }
    \label{fig:darkstate_twores}
\end{figure}

To illustrate in the simplest setting how the interplay of the different system parameters gives rise to a dark state, let us first consider the minimal subsystem formed by the two mutually coupled resonators $\Omega_2$ and $\Omega_3$. Two resonators coupled to a common transmission waveguide already capture the essential interference mechanism underlying the formation of a stationary dark state. Given the jump operators $\hat{L}_{\pm}$ for both propagation directions, the dark state condition that suppresses the radiation into the common feedline for its stationary states reads $\hat{L}_\pm \ket{\psi_\pm} = 0$, 
\begin{equation}
  \left[\sqrt{\kappa_3}\, \hat{a}_3 + e^{\pm i\Theta}\sqrt{\kappa_2}\,\hat{a}_2\right]\ket{\tilde{\psi}_{\pm}}=0 ,
  \label{eq:twores_condition_maintext}
\end{equation}
where $\Theta$ is the total propagation phase between input and output fields. Solving these equations reveals two coupling regimes under which darkness can be induced (\textcolor{black}{see Appendix~\ref{sec:app_darkstate_twores}}).
In the case of zero cavity-cavity interaction, \(\zeta = 0\), the system has to satisfy the special case of exact degeneracy, \(\Omega_2 = \Omega_3\), together with \(\kappa_2 = \kappa_3\)\,\cite{Propp22, PhysRevA.79.042302}. Otherwise, a finite resonator-resonator interaction, \(\zeta \neq 0\), is essential for the emergence of a stationary dark state. In this case, the amplitude of the couplings $\kappa_2$ and $\kappa_3$ must verify
\begin{equation}
    \tan\left(\frac{\varphi}{2} \right) =  \sqrt{ \frac{ \vert \kappa_2 \vert }{\vert \kappa_3 \vert} } ,
    \label{eq:magnitude_twores}
\end{equation}
while the phases must satisfy
\begin{equation}
    \pm \Theta + (\phi_3 - \phi_2)/2 = \pm (2n-1)\pi, \quad n \in \mathds{Z_+}
    \label{eq:phases_twores}
\end{equation}
for the left and right propagating channel. For perfect couplings $\kappa_i = \vert \kappa_i\vert$, the dark state condition is automatically fulfilled for both propagation channels, as long as $\vert \Theta \vert = (2n-1)\pi$. However, the imperfections of the device not only shift the destructive-interference condition, but they also generally turn a complete dark state (i.e., dark to both transmission and reflection) into a directionally dark state (dark to either transmission or reflection) if the following conditions are not further satisfied: $\Theta = k\pi$, $\phi_3 - \phi_2 = 2l\pi$, with $k + l \in \mathds{Z}_{\mathrm{odd}}$. \\
Let us define 
\begin{equation}
    f_{\text{2r}}^{\pm}(\vec{v}) = \sqrt{\kappa_3}\, \sin\left(\frac{\varphi}{2}\right) + e^{\pm i\Theta} \sqrt{\kappa_2} \cos\left( \frac{\varphi}{2}\right)
\end{equation}
as the left-hand side of Eq. \eqref{eq:twores_condition_maintext}, where $\vec{v} = (\kappa_2,\kappa_3,\varphi,\Theta)$ are all the parameters needed for evaluation. In Fig.\,\ref{fig:darkstate_twores} we show that the total interference phase of Eq. \eqref{eq:phases_twores} controls whether the two emitted fields cancel in both propagation directions simultaneously, producing a fully dark state, or only in one direction. As the right- and left-moving channel represent orthogonal decay environments, the operational point with the smallest total radiative decay into the feedline can be obtained through $F_{\text{2r}}(\vec{v}) = \vert f^{+}_{{\text{2r}}}(\vec{v})\vert^2 +  \vert f^{-}_{\text{2r}}(\vec{v})\vert^2$. 
Fig.\,\ref{fig:darkstate_twores} shows the dark state condition $f_{\text{2r}}^\pm(\vec{v})/\mathrm{max}[f^\pm_{\text{2r}}]$ for different phases $\{\phi_2,\phi_3\}$, as a function of $\Theta$, given that the condition Eq. \eqref{eq:magnitude_twores} is fulfilled. Panel \ref{fig:darkstate_twores}(a) shows a perfect dark state for $\Theta = 0$, as the phases $\phi_i$ are chosen as $\vert \phi_3 - \phi_2\vert = 2\pi$. Notably, for the parameters shown in Fig. \ref{fig:darkstate_twores}(b), the change in the directional decays as a function of $\Theta$, and the presence of directional dark states for specific $\Theta$ values, does not change the total decay, which remains constant, while for panel \ref{fig:darkstate_twores}(c), the operational point with the smallest environmental decay happens at $\Theta = 0$, which does not coincide with the presence of directional dark states.\\

The present results extend the concept of a dark state beyond the strictly symmetric configuration and incorporate complex resonator-transmission couplings.
This generalization enables more flexible device design and enhances the robustness of the dark-state condition. In practical implementations, nominally identical resonators often exhibit deviations such as unequal linewidths $\kappa_2 \neq \kappa_3$, variations in local coupling to ancillary subsystems, or small fabrication-induced detunings. Under these circumstances, the exact symmetry point becomes fragile.
A tunable $\zeta$ or cavity detuning $\Omega_-$ can compensate for these asymmetries by adjusting the mixing angle $\varphi$ \cite{VallSanclemente2023, PhysRevApplied.14.044040, PhysRevApplied.19.034021}. This tunability preserves the existence of a dark state even when the system departs from perfect symmetry.
Furthermore, even for $\Omega_2 = \Omega_3 \equiv \Omega$, a non-zero direct coupling lifts the degeneracy of the hybridized modes and results in a splitting of the two hybridized modes $\tilde{\Omega}_\pm$ by $2\zeta$.
Consequently, once one of these hybridized modes is tuned to satisfy the dark-state condition, the dark and bright sectors remain spectrally separated. This separation is advantageous for selective addressing and for improving the detectability of the dark mode.\\

From a different perspective, the finite resonator-resonator coupling provides a coherent channel for hybridization and excitation transfer, and tuning $\zeta$ and $\Omega_-$ can be exploited to access dark-mode configurations without requiring additional driving fields \cite{PhysRevA.84.061805, Zanner2022}. In our setup, the spin ensemble coupled to one of the resonators acts as the control knob to tune in the dark state condition. Of course, the presence of the electronic system changes the composition of the eigenstates, such that a new dark-mode must be redefined for the tripartite system. But once the needed operational point is found, the advantage now is that by externally modifying the matter energy splitting, one enables tunable conversion between the dark and bright configurations for a given eigenstate: by changing the spin splitting via the Zeeman interaction with a magnetic field, the spin mode can be tuned in and out of the dark state condition. Then, for a given target eigenstate, a bright configuration couples to the waveguide and can be efficiently used for loading and readout, whereas the dark one, being decoupled from radiative channels, can store the excitation for longer times. Lastly, for $\zeta = 0$ and a finite detuning $\Omega_2 \neq \Omega_3$, the Hamiltonian eigenstates are just the bare local cavity excitations, $\ket{1,0}$ and $\ket{0,1}$, and none of them satisfy $\hat{L}_p\ket{n_2, n_3} = 0$. This is expected, as a bare local cavity excitation cannot be dark. \\

\begin{figure}
    \includegraphics[]{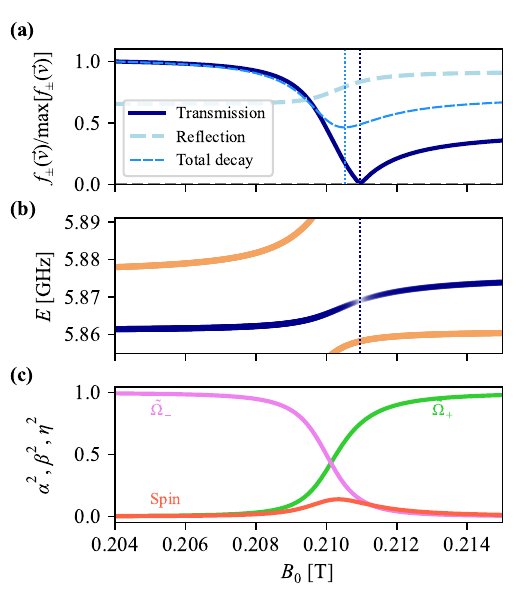}
    \caption{\label{fig:darkstate_tripartite} \textbf{Dark state in the tripartite system.} \textbf{(a)} Dark state condition for the central eigenstate of the Hamiltonian in Eq. \eqref{eq:hamiltonian_spin_hybrized}, evaluated through $f_{\pm}(\vec{v}) / \max(f_{\pm}(\vec{v}))$, using the parameters $\vec{v} = (\kappa_2,\kappa_3,\alpha,\beta, \varphi, \theta)$ obtained from the fitting in Table \ref{tab:spin_hamiltonian}. The observed dark state is only dark to transmission. \textbf{(b)} Energy spectrum for the same fitting parameters. The opacity of the lines corresponds to the value of $|f_+(\vec{v})| / \max(|f_+(\vec{v})|)$, which positions a dark state for the central branch at $B_0 \approx 211$ mT. \textbf{(c)} Composition of the central eigenstate of the system as a function of $B_0$.}
\end{figure}

Having illustrated in a simplified setting how the interplay of the different parameters determines the formation of a dark state, we now turn to the full experimental setup. We determine the parameter regimes in which it can support a stationary dark state and compare the resulting predictions with the experimental measurements presented above. From the Hamiltonian in Eq. \eqref{eq:hamiltonian_spin_hybrized}, a general eigenstate can be written as
\begin{equation}
    \ket{\Psi} = \alpha \ket{\downarrow, \tilde{\psi}_+} + \beta \ket{\downarrow, \tilde{\psi}_-} + \eta \ket{\uparrow,0,0},
\end{equation}
where $\alpha,\beta,\eta$ are probability amplitudes. They will be evaluated numerically for the purposes of this section, though the eigenvalue problem of Eq. \eqref{eq:hamiltonian_spin_hybrized} can be solved analytically as well \cite{mao2025genuinetripartitestrongcoupling}. In our system composed of the spin ensemble and two resonators, the dark-state condition reads,
\begin{align}
    \sqrt{\kappa_3}& \left[\alpha \sin\left( \frac{\varphi}{2}\right) + \beta \cos\left(\frac{\varphi}{2} \right)\right] \nonumber \\
    & +e^{\pm i\Theta}\sqrt{\kappa_2}\left[ \alpha \cos\left(\frac{\varphi}{2} \right) - \beta \sin\left( \frac{\varphi}{2}\right) \right] = 0,
    \label{eq:darkstate_con_3}
\end{align}
where we have also used Eqs.\,\eqref{eq:eigenstates_plus} and  \eqref{eq:eigenstates_minus}. Notice that resonator 1 does not enter the dark-state problem because it is far detuned and does not coherently couple to the rest of the elements in the circuit, but we include the total traveling distance between the input and the output in the phase $\Theta$. By calling $f_{\pm}(\vec{v})$ the left-hand-side of Eq.\,\eqref{eq:darkstate_con_3} with $\pm \Theta$, and $\vec{v}$ containing all the relevant parameters to be used in the evaluation, $\vec{v} = (\kappa_2,\kappa_3,\alpha,\beta,\varphi, \theta) $, we can now numerically evaluate $\vert f_+(\vec{v})\vert / \max(\vert f_+(\vec{v})\vert) = 0$ to test the presence or absence of a dark state within the parameter region of the experiment, as shown in Fig. \ref{fig2}. Our numerical results, shown in Fig. \ref{fig:darkstate_tripartite}, confirm the presence of a transmission-dark state in the central mode for the fitted parameters of Table \ref{tab:spin_hamiltonian} and $\Theta \approx 2.2$\,rad. This value is in excellent agreement with the estimated phases $\theta_{2,3}$, since $\Theta=\theta_2+\theta_3$. However, device imperfections, encoded in the complex $\kappa_i$, prevent the state from being simultaneously dark in reflection. Nevertheless, we identify a configuration exhibiting a subradiant mode, for which the total decay into the transmission line is reduced through the combined interference of the left- and right-propagating decay channels. This behavior is shown by the dashed medium-blue curve in Fig.\,\ref{fig:darkstate_tripartite}(a). For the experimental parameters, the other two modes do not support a dark state, except in the limiting case where the mode becomes a pure spin eigenstate (see Appendix \ref{app:composition_other_eigenstates}).\\

\section{Conclusion and outlook}

In summary, we have characterized a multi-element hybrid system composed of three superconducting microwave resonators coupled to a common transmission line, and a spin ensemble interacting directly with one of the resonators. The transmission spectra of our hybrid system reveal a distinct multi‑mode hybridization mediated by direct cavity–cavity interaction. The observed pattern of avoided crossings cannot be interpreted as the coupling of the spin ensemble to an isolated bare resonator; instead, the direct interaction between resonators 2 and 3 produces two hybridized photonic modes, both of which acquire a finite coupling to the spin ensemble. The inter-resonator coupling and cavity detuning redistribute the local light-matter interaction between these two branches, leading to mode-dependent spin-photon coupling strengths, linewidths, and cooperativities. For the parameters extracted from the experiment, both hybridized photonic modes satisfy the corresponding strong-coupling condition, demonstrating multi-mode strong coupling within the spin-resonator-resonator subsystem.

In addition, we identify a pronounced suppression of the central transmission feature as a signature of destructive interference between the fields radiated by the different components of the hybridized mode, which highlights their coherent nature. Using the jump operators associated with the two propagation directions of the common transmission line, we derive the conditions under which a stationary eigenstate of the hybrid system becomes dark to transmission, reflection, or both channels simultaneously. The spin resonance provides an external control parameter that changes the composition of the hybridized eigenstates and thereby tunes a given branch between bright and dark configurations. Together with the resonator detuning and inter-resonator coupling, this provides a route to mode-selective loading, readout, and radiative protection without requiring additional control drives.

Consequently, the demonstrated role of the dissipative bath mode suggests a promising route towards engineered dissipation in multi‑cavity quantum circuits. By tailoring the spectral properties and coupling strengths of such bath modes, it may become possible to selectively enhance or suppress specific hybridized excitations, thereby enabling controlled formation of dark states and other interference‑based phenomena. This approach offers a potential platform for generating non-classical states of collective spin excitations through dissipation engineering. In particular, structured bath environments could be exploited to realize protected subspaces, long‑lived coherent superpositions, or dissipatively prepared entangled states in multi‑mode cavity architectures. These capabilities would provide valuable tools for quantum information processing, quantum simulation, and the exploration of many‑body physics in hybrid quantum systems.

\section*{Acknowledgments}
We acknowledge financial support by the Deutsche Forschungsgemeinschaft (DFG, German Research Foundation) via the Transregio ConQuMat TRR 360-C7 (Project ID 492547816) and via Germany’s Excellence Strategy EXC-2111-390814868. This research is part of the Munich Quantum Valley (lighthouse project IQ-Sense), which is supported by the Bavarian state government with funds from the Hightech Agenda Bayern Plus. BPG acknowledges the Alexander von Humboldt Institution through the postdoctoral fellowship program.


\section*{Author contribution statement}
PO and HH conceived the experiment. PO and AD performed the experiments. BPG developed the theoretical model with input from MB and FP. PO performed the experimental data analysis. PO, BPG, MB and HH wrote the manuscript with help from all authors.

\appendix

\section{Experimental setup \label{app: Appendix Experimental Setup}}
Our microwave circuit consists of three planar, superconducting lumped-element resonators that are coupled to a transmission line waveguide in a hanger-type geometry. 
For the fabrication of the resonator device, we use a high-resistivity silicon substrate, on which we deposit a $\SI{150}{\nano\meter}$ thin niobium film and pattern the circuit design with optical lithography and reactive ion etching techniques. 
The spin ensemble is provided by DPPH, which is a stable free radical used as a standard for electron spin resonance spectroscopy\,\cite{yordanov1996our}. Each DPPH radical hosts an electron spin $S=1/2$ with an isotropic $g=2.0036$ factor at room temperature\,\cite{weil2007electron}. 
We note that the $g$-factor shows a strong temperature dependence reaching values up to 2.1 for millikelvin temperatures\,\cite{voesch2015dpph}. However, we here focus on the multi-mode coupling physics and therefore adjust the field strength at the position of our spin sample using a constant magnetic field offset to reference the resonance field to a $g$-factor of $g=2.004$\,\cite{roca2025waveguide}.
The spin ensemble is exclusively placed on the third resonator in a flip-chip geometry and we ensure no physical contact with the other resonators or the transmission line waveguide. 
The composed multi-cavity spin hybrid system is probed using a vector network analyzer within a wet cryostat at a temperature of $\SI{4}{\kelvin}$. To thermalize the microwave coaxial line and improve the signal-to-noise ratio, we use a cryogenic attenuator for the input line and a cryogenic HEMT amplifier at the output line, respectively.
The static magnetic field $B_0$ is provided by a superconducting solenoid magnet. The field is applied in-plane to the Niobium thin-film and oriented in parallel to the transmission line. 

\section{Background correction and magnetic field calibration}
\label{Appendix: background correction}
\begin{figure}[t]
    \centering
    \includegraphics[width=1\linewidth]{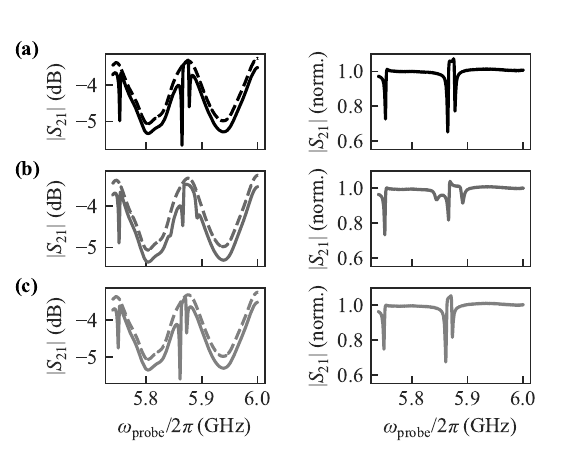}
    \caption{\textbf{Background correction.} Dashed lines: Frequency-dependent microwave transmission measured at \SI{1}{\tesla}. Comparison of the raw measured microwave transmission (left panels) and corrected microwave transmission (right panels) for magnetic field strengths of (a) $B_0=\SI{198}{\milli\tesla}$, (b) $B_0=\SI{209}{\milli\tesla}$ and (c) $B_0=\SI{223}{\milli\tesla}$.}
    \label{fig appendix: background correction}
\end{figure}

During the measurements of the cwESR spectrum, the measured transmission parameter is subject to a complex microwave background based on imperfections in the experimental setup
such as impedance mismatches, unwanted frequency-dependent microwave reflections, electronic delay, or finite cable length. In order to correct for effects that are independent on the applied magnetic field, we record a reference spectrum at an elevated magnetic field of $B_0 = \SI{1}{\tesla}$. For the superconducting microwave resonators, this magnetic field strength is far above the critical field and they are no longer detectable\,\cite{muller2022magnetic}. 
We then correct our measured cwESR spectrum with this reference measurement: 
\begin{align}
    |S_{21}'| = \Bigg|\frac{S_{\text{21, measured}}}{S_{\text{21, 1T}}}\Bigg| \text{.}
\end{align}
Moreover, we observe a reduction in the microwave transmission as a function of frequency. Over the limited frequency interval considered here, this trend can be well approximated by a linear dependence. At the lowest applied magnetic field, we extract this linear background by fitting the frequency‑dependent transmission and subsequently use it to correct the measured cw‑ESR spectrum: 
\begin{align}
    |S_{21}| = \Bigg| \frac{S_{21}'}{m \cdot \omega_{\text{probe}}/(2\pi)-n} \Bigg| \text{,}
\end{align}
where $m$ describes the slope of the linear reduction and $n$ for an offset of the microwave transmission. Figure\,\ref{fig appendix: background correction} shows a total of three examples for the frequency-dependent background correction. 

In addition to a complex microwave background, we are also subject to off-set magnetic fields at the sample position within our experimental setup. Those can origin from an unwanted electromagnetic environment in our lab or the earth magnetic field, since our sample is not shielded within the cryostat. To correct for the offset magnetic field and compare the measured cwESR spectrum with the modeled data, we rely on the ESR transition of our DPPH sample. Assuming the known g factor of DPPH, allows us to calibrate the magnetic field strength at the sample position\,\cite{velluire2023hybrid}. We determine the off-set magnetic field in our experiments to be $\Delta B_0=B_{\text{0}}-B_{\text{0, set}}=\SI{7.223}{\milli\tesla}$. 

\section{Input-output theory}
\label{Appendix: details of theory}
As mentioned in the main text, we assume a total of three resonators that are directly coupled to a transmission line waveguide, modeled as the Hamiltonian shown in Eq.\,\eqref{eq: Hamiltonian cav-tl}. The Heisenberg equations of motion (EOMs) $d(\cdot)/dt = -i[(\cdot),\hat{\mathcal{H}}]$ for each of the three individual resonators yield
\begin{align}
    \frac{d \hat{a}_1}{dt}  & = -i \Omega_1 \hat{a}_1 - \sum_{p=\pm} \sqrt{\frac{\kappa_{1}}{2\pi}}\int_0^{\infty} d\nu  \, \hat{d}_{p,1}(\nu) \mathrm{,}
    \label{eq:EOM_cav_1}
\end{align}

\begin{align}
\begin{split}
    \frac{d \hat{a}_2}{dt}  = &-i \Omega_2 \hat{a}_2 - \sqrt{\frac{\kappa_{2}}{2\pi}}\sum_{p=\pm} \int_0^{\infty} d\nu \,  \hat{d}_{p,2}(\nu) \\
    &  - i \zeta_{23}\hat{a}_3 \mathrm{,}
    \label{eq:EOM_cav_2}
\end{split}
\end{align}

\begin{align}
\begin{split}
    \frac{d\hat{a}_3}{dt}  = &-i \Omega_3 \hat{a}_3 - \sqrt{\frac{\kappa_{3}}{2\pi}}\sum_{p=\pm}  \int_0^{\infty} d\nu  \, \hat{d}_{p,3}(\nu)\\
    &  - i \zeta_{23}\hat{a}_2 - ig\hat{\sigma}_- \mathrm{.}
    \label{eq:EOM_cav_3}
\end{split}
\end{align}

Similarly, the EOM for the waveguide modes gives
\begin{align}
    \frac{d }{dt} \hat{d}_{p,j}(\nu) = -i \nu \hat{d}_{p,j}(\nu) + \sqrt{\frac{\kappa_{j}}{2\pi}} \hat{a}_j \mathrm{.}
    \label{eq:waveguide_mode_EOM}
\end{align}
%
%
Eq. \eqref{eq:waveguide_mode_EOM} can be easily integrated to give
\begin{equation}
    \hat{d}_{p,j}(\nu) = \hat{d}_{p,j}(\nu,t_0) e^{-i\nu\left(t-t_{0}\right)} +   \sqrt{\frac{\kappa_{j}}{2\pi}} \int_{t_{0}}^{t}dt^{\prime} \, \hat{a}_j(t)  e^{-i\nu\left(t-t^{\prime}\right)}\mathrm{,}
\end{equation}
expressed in terms of the initial condition $t_0 < t$. Substituting this result back into Eqs. \eqref{eq:EOM_cav_1}, \eqref{eq:EOM_cav_2} and \eqref{eq:EOM_cav_3}, we can define the input fields for each resonator as 
\begin{align}
    \hat{d}^{\text{in}}_{p,j}(t) = \frac{1}{\sqrt{2\pi}} \int_0^{\infty} d\nu\,\hat{d}_{p,j}(\nu, t_0) \,e^{-i\nu(t-t_0)}.
\end{align}
Together with the property $\int_{t_0}^{t} f(t^\prime)\delta(t-t^\prime) = f(t)/2$ and after moving to frequency space with a Fourier transform, the EOM for each resonator can be written as
\begin{subequations}
    \begin{align}
        \left[ -i(\omega - \Omega_1) +  \frac{\kappa_{\mathrm{cav},1}}{2} \right]\hat{a}_1  = &  - \sqrt{\kappa_{1}} \, (\hat{d}^{\mathrm{in}}_{+,1} + \hat{d}^{\mathrm{in}}_{-,1}), \label{eq:eom_1}\\
    \left[ -i(\omega - \Omega_2) + \frac{\kappa_{\mathrm{cav},2}}{2} \right]\hat{a}_2 = & - \sqrt{\kappa_{2}} \, (\hat{d}^{\mathrm{in}}_{+,2} + \hat{d}^{\mathrm{in}}_{-,2}) \nonumber \\
    &  -i\zeta_{} \hat{a}_3, \\
    \left[ -i(\omega - \Omega_3) + \frac{\kappa_{\mathrm{cav},3}}{2} \right]\hat{a}_3  = & - \sqrt{\kappa_{3}} \, (\hat{d}^{\mathrm{in}}_{+,3} + \hat{d}^{\mathrm{in}}_{-,3}) -i\zeta \hat{a}_2 \nonumber \\
    & -\left( \frac{ig^2}{\omega - \Delta +i\frac{\gamma}{2}} \right) \hat{a}_3. \label{eq:eom_3}
    \end{align}
\end{subequations}

We have defined the total cavity linewidth as $\kappa_{\mathrm{cav},j}/2 = (\kappa_{\mathrm{int},j} + 2\vert \kappa_{j} \vert )/2$, which comprises the equal contributions of each port $\pm$, together with the intrinsic decay rate $\kappa_{\mathrm{int},j}/2$.\\

Upon integrating Eq.~\eqref{eq:waveguide_mode_EOM} with the boundary condition $t_1 > t$, the EOMs can be recast in terms of the output fields,
\begin{align}
    \hat{d}^{\text{out}}_{p,j}(t) = \frac{1}{\sqrt{2\pi}} \int_0^{\infty} d\nu\,\hat{d}_{p,j}(\nu, t_1) \,e^{-i\nu(t-t_1)}.
\end{align}
and obtain the input-output relations for each resonator in the setup,
\begin{equation}
    \hat{d}^{\text{out}}_{p,j} = \sqrt{\kappa_j} \, \hat{a}_j + \hat{d}^{\text{in}}_{p,j}.
    \label{eq:input_output}
\end{equation}
Here, the right-traveling input field $\hat{d}^{\,\mathrm{in}}_{+,j}$ brings the information of the previous $m<j$ resonators, while the left-traveling field $\hat{d}^{\,\mathrm{in}}_{-,j}$ contains information related to the $m>j$ resonators to follow. Using Eq.~\eqref{eq:input_output} together with the relations Eqs.~\eqref{eq:input_output_withphase}, we can write
\begin{equation}
   \hat{d}^{\,\mathrm{in}}_{+,j} = e^{i\theta_j}\hat{d}^{\, \mathrm{out}}_{+,j-1} \\
    =  e^{i\theta_j} \sqrt{\kappa_{j-1}} \,\hat{a}_{j-1} + e^{i\theta_j} \hat{d}^{\, \mathrm{in}}_{+,j-1}.
\end{equation}
From this result, we learn that the common coupling to the transmission line also generates an indirect interaction between each pair of resonators $n$ and $m$, of the order $\sqrt{\kappa_n \kappa_{m}}$ \cite{Propp22, xiao2010asymmetric}.

\section{Test case: two mutually coupled resonators}

Let us analyze the case of two mutually coupled resonators, radiating into a common transmission line. For direct comparison with our experimental setup, we keep the labeling $\Omega_2$ and $\Omega_3$, without including resonator 1. 

\subsection{Decay rates for hybridized photonic modes \label{sec:app_decaylength_twores}}

We first write the EOM for the interacting resonators, following the same procedure as for Eqs.~\eqref{eq:eom_1}-\eqref{eq:eom_3}. We get

\begin{equation}
    \left[ -i(\omega - \Omega_2) +  \frac{\kappa_{\text{cav},2}}{2}\right]\hat{a}_2 =   - \sqrt{\kappa_{2}} \, (\hat{d}^{\mathrm{in}}_{+,2} + \hat{d}^{\mathrm{in}}_{-,2}) - i\zeta \hat{a}_3,
\end{equation}

and 
\begin{equation}
    \left[ -i(\omega - \Omega_3) + \frac{\kappa_{\text{cav},3}}{2}\right]\hat{a}_3 = - \sqrt{\kappa_{3}} \, (\hat{d}^{\mathrm{in}}_{+,3} + \hat{d}^{\mathrm{in}}_{-,3}) - i\zeta \hat{a}_2. 
\end{equation}

The input-output relations for this system,
\begin{equation}
    \hat{d}^{\text{out}}_{+,j} = \sqrt{\kappa_j}\, \hat{a}_j + \hat{d}^{\text{in}}_{+,j},\quad j=2,3
\end{equation}
together with the phases induced due to the propagation distance between the two, 
\begin{equation}
    \hat{d}^\mathrm{in}_{+,3} = e^{i\theta_3}\hat{d}^\mathrm{out}_{+,2},\quad \hat{d}^\mathrm{in}_{-,2} = e^{i\theta_3}\hat{d}^\mathrm{out}_{-,3},
\end{equation}
let us write the output, measurable field in terms of the input field,
\begin{eqnarray}
    \hat{d}^{\text{out}}_{+,3} & = & \sqrt{\kappa_3}\, \hat{a}_3 + \hat{d}^{\text{in}}_{+,3} \nonumber \\
    & = & \sqrt{\kappa_3}\, \hat{a}_3 + e^{i\theta_3}\hat{d}^{\text{out}}_{+,2} \nonumber  \\
    & = &  \sqrt{\kappa_3}\, \hat{a}_3 + e^{i\theta_3}\sqrt{\kappa_2}\,\hat{a}_2 + e^{i\theta_3}\hat{d}^{\text{in}}_{+,2}.
    \label{eq:out_vs_in_2res}
\end{eqnarray}
The transmission coefficient, $S^{\text{2r}}_{21}(\omega) = \braket{\hat{d}^{\mathrm{out}}_{+,3}}/\braket{\hat{d}^{\mathrm{in}}_{+,2}}$, has a simple analytical expression,
\begin{equation}
    S^{\text{2r}}_{21}(\omega) = e^{i\theta_3}\frac{[ \chi_2(\omega) - \kappa_2][\chi_3(\omega) - \kappa_3] + \zeta^2 + 2 \zeta K \sin(\theta_3)}{\chi_2(\omega)\chi_3(\omega) + G_{\theta_3}^2},
\end{equation}
where we have defined the cavity susceptibilities as $\chi_i(\omega) = \kappa_{\mathrm{cav},i}/2 - i(\omega - \Omega_i)$, and the effective complex inter-resonator coupling $G_{\theta_3} = \zeta - iKe^{i\theta_3}$, with $K = \sqrt{\kappa_2 \kappa_3}$. The term $Ke^{i\theta_3}$ contributing to $G_{\theta_3}$ can be interpreted as the transmission-mediated effective interaction between resonators, added to their coherent coupling $\zeta$.\\

From the zeros of the denominator
\begin{equation}
D(\omega) = \chi_2(\omega)\chi_3(\omega) + [\zeta - iKe^{i\theta_3}]^2,
\end{equation}
we find two complex poles of the form $\tilde{\omega}_\pm = \tilde{\Omega}^\prime_\pm - \frac{i}{2}\tilde{\kappa}_\pm$. On the one hand, the real part $\tilde{\Omega}^\prime_\pm = \text{Re}[\tilde{\omega}_\pm]$ does not entirely correspond to the previously obtained eigenenergies $\tilde{\Omega}_{\pm}$ of Eq. \eqref{eq:twores_eig}. Instead, it additionally captures the effect of the transmission-line-mediated resonator-resonator interaction \cite{PhysRevE.64.036213, Davy2018},
\begin{align}
    \tilde{\Omega}^\prime_{\pm}  = &\frac{\Omega_2 + \Omega_3}{2} \nonumber  \\
     & \pm \frac{1}{2} \mathrm{Re} \left( \left[  \Omega_-  - \frac{i(\kappa_{\text{cav},2} - \kappa_{\text{cav},3})}{2} \right]^2   + 4G_{\theta_3}^2 \right)^{1/2}\mathrm{.}
    \label{eq:exact_modes_with_diss}
\end{align}
On the other hand, the imaginary part of the poles $\mathrm{Im}\left[ \tilde{\omega}_{\pm}\right] =-\tilde{\kappa}_{\pm}/2$ gives their corresponding effective linewidth, namely 
\begin{align}
    \tilde{\kappa}_\pm  = & \, \dfrac{\kappa_{\text{cav},2} + \kappa_{\text{cav},3}}{2} \nonumber \\
    &  \mp \text{Im}\left(\left[  \Omega_-  - \frac{i(\kappa_{\text{cav},2} - \kappa_{\text{cav},3})}{2}\right]^2+4G_{\theta_3}^2 \right)^{1/2},
    \label{eq:linewidth_Exact}
\end{align}
with  the effective complex inter-resonator coupling $G_{\theta_3} = \zeta - iKe^{i\theta_3}$, and $K = \sqrt{\kappa_2 \kappa_3}$. Let us mention that the numerator $N(\omega)$ can produce Fano distortions, shift the apparent spectral maxima, or suppress one of the two modes if there is a strong frequency dependence, so the final visibility of the pole in the measurement is not fully determined by the denominator structure only.\\

First, we can understand this result in a simpler case. In the scenario where $\Omega_2 =  \Omega_3 \equiv \Omega$, $\kappa_{\mathrm{cav},2} = \kappa_{\mathrm{cav},3} \equiv \kappa_{\mathrm{cav}}$, and $\{ \kappa_2, \kappa_3 \} \in \mathds{R} $, the poles reduce to 
\begin{equation}
    \tilde{\omega}_{\pm} = \Omega \pm \left[\zeta + K\sin(\theta_3)\right]-\frac{i}{2}\left[ \kappa_{\text{cav}} \pm 2 K \cos(\theta_3)\right]
    \label{eq:simplified_normal_loss}
\end{equation}
and therefore the linewidths are given by $\tilde{\kappa}_{\pm} = \kappa_{\mathrm{cav}} \pm 2 K \cos(\theta_3)$. Interestingly, the propagation phase plays an active role in the effective description, entering in $G_{\theta_3}$ and producing two effects: first, a shift in the resonator-resonator coupling, $\zeta \rightarrow \zeta + K \sin(\theta_3)$, which contributes to coherent level repulsion, and second, a modulation of the collective radiative damping term $K \cos(\theta_3)$. To zeroth order in $\kappa_{\mathrm{cav}}$, one can identify $\tilde{\Omega}_\pm \approx \tilde{\omega}_\pm$. \\
Now, for the standard case of $\theta_3 = 0$ and $\kappa_{2}= \kappa_3 \equiv \kappa$, we obtain the usual bright/dark splitting of decay rates \cite{Trebbia2022} of $\tilde{\kappa}_{\pm} = \kappa_{\mathrm{cav}} \pm 2\kappa$. Recalling that $\kappa_{\mathrm{cav}} = 2\kappa+\kappa_\mathrm{int}$, we can write
\begin{equation}
    \tilde{\kappa}_+ = 4\kappa  + \kappa_{\text{int}},\quad  \tilde{\kappa}_- = \kappa_{\mathrm{int}}.
\end{equation}
As expected, the antisymmetric mode is dark with respect to the common waveguide, and its residual linewidth is only the intrinsic loss.\\

Notice that the solution for the poles of $S^{\text{2r}}_{21}(\omega)$ corresponds to the eigenvalues of a non-Hermitian Hamiltonian \cite{PhysRevLett.104.153601, PhysRevResearch.3.023093, PhysRevB.96.235434}, such that 
\begin{equation}
    \hat{H}_{\text{eff}} = \left( \begin{array}{cc}
    \Omega_2 - i\kappa_{\text{cav},2}/2 & G_{\theta_3} \\
    G_{\theta_3} & \Omega_3 - i\kappa_{\text{cav},3}/2 
    \end{array} \right).
\end{equation}
In the hybridized basis of the two resonators, given by Eqs. \eqref{eq:eigenstates_plus} and \eqref{eq:eigenstates_minus}, the Hamiltonian takes the form,
\begin{equation}
    \hat{H}^{\text{R}}_{\mathrm{eff}} = \left( \begin{array}{cc}
    \tilde{\Omega}^{\text{R}}_+ - i\Gamma_+/2 & \Gamma_{+-} \\
    \Gamma_{-+} & \tilde{\Omega}^{\text{R}}_- - i\Gamma_-/2 
    \end{array} \right).
\end{equation}
As expected, the eigenvectors of the coherent sector are not eigenvectors of the complete, dissipative Hamiltonian. To leading order in the damping terms, one can assume the coherent modes to be
\begin{equation}
    \tilde{\Omega}_{\pm}^{\mathrm{R}} = \tilde{\Omega}_\pm \pm  \sqrt{\vert \kappa_2 \kappa_3 \vert } \sin\left(\theta_3 - \frac{\phi_2 + \phi_3}{2}\right)\sin(\phi),
\end{equation}
with linewidths of
\begin{subequations}
    \begin{align}
        \Gamma_+   = & \kappa_{\mathrm{cav},2}\cos^2\left(\frac{\varphi}{2}\right) + \kappa_{\mathrm{cav},3}\sin^2\left(\frac{\varphi}{2}\right) \nonumber \\
    &  + 2\sqrt{\vert \kappa_2 \kappa_3 \vert }\cos\left(\theta_3 - \frac{\phi_2 + \phi_3}{2} \right)\sin\left( \varphi \right), \label{eq:linewidthplus}\\
    \Gamma_-   = & \kappa_{\mathrm{cav},2}\sin^2\left(\frac{\varphi}{2}\right) + \kappa_{\mathrm{cav},3}\cos^2\left(\frac{\varphi}{2}\right) \nonumber \\
    &  - 2\sqrt{\vert \kappa_2 \kappa_3 \vert }\cos\left(\theta_3 - \frac{\phi_2 + \phi_3}{2} \right)\sin\left( \varphi \right).\label{eq:linewidthminus}
    \end{align}
\end{subequations}
Here we have made explicit the complex phase of each $\kappa_i = \vert \kappa_i \vert e^{-i\phi_i}$, to correctly identify the real and imaginary parts of the diagonal terms in the Hamiltonian. These equations correspond to Eq. \eqref{eq:linewidth_resonantmodes} in the main text. The off-diagonal elements $\Gamma_{+-} = \Gamma_{-+}$ provide an additional coherent and dissipative coupling between the modes, mediated by the common transmission line, 
\begin{eqnarray}
    \Gamma_{+-} & = &  K \sin(\theta_3)\cos(\varphi)-i K\cos(\theta_3) \cos(\varphi) \nonumber  \\
    && \hspace{5pt} - \frac{i}{4}(\kappa_{\mathrm{cav},3} - \kappa_{\mathrm{cav},2})\sin(\varphi).
\end{eqnarray}
If neglected, one can directly identify $\tilde{\kappa}_{\pm} \approx \Gamma_{\pm}$. Let us note that in the case of identical resonators, $\Omega_2 = \Omega_3$ and $\kappa_{\mathrm{cav},2} = \kappa_{\mathrm{cav},3}$, we obtain $\Gamma_{+-} = 0$ and the rotated modes are true normal modes.\\

\begin{figure}
    \centering
    \includegraphics{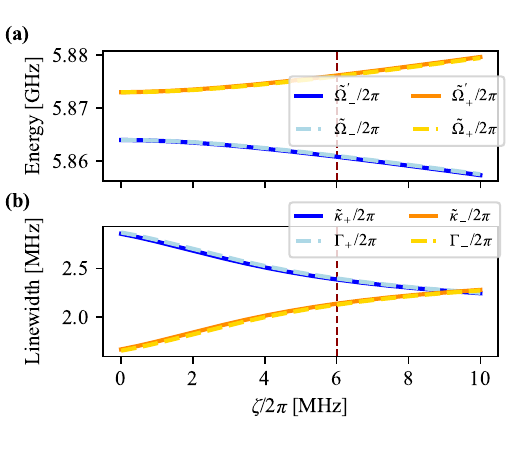}
    \caption{\textbf{Comparison of mode frequency and linewidth obtained from different methods.} \textbf{(a)} Energy mode $\tilde{\Omega}_\pm^\prime = \mathrm{Re}(\tilde{\omega}_\pm)$ from Eq. \eqref{eq:exact_modes_with_diss}, versus $\tilde{\Omega}_{\pm}$, obtained in Eq. \eqref{eq:twores_eig} from diagonalizing the Hermitian Hamiltonian matrix. \textbf{(b)} Effective linewidth of the hybridized modes, obtained from Eq. \eqref{eq:linewidth_Exact} (exact result), and from Eqs. \eqref{eq:linewidthplus} and \eqref{eq:linewidthminus}. The parameters used are those of the experiment, as fitted in Tab. \ref{tab:spin_hamiltonian}, as a function of $\zeta$. The vertical, dashed line indicates the fitted value for $\zeta$.}
    \label{fig:linewidths}
\end{figure}

In Fig. \ref{fig:linewidths}(a) we compare the mode energies obtained in the main text, Eq. \eqref{eq:twores_eig}, from the diagonalization of the coherent Hamiltonian, with the ones obtained from the poles of $\vert S_{21}(\omega)\vert$ shown in Eq. \eqref{eq:exact_modes_with_diss}. The parameters used are the fitted values of Table \ref{tab:spin_hamiltonian}. They perfectly agree over a reasonable range of $\zeta$ values,  with the dashed, vertical line corresponding to the value for the experiment. In panel \ref{fig:linewidths}(b), we plot the total linewidth in Eq. \eqref{eq:linewidth_Exact} and compare it to the result in Eqs. \eqref{eq:linewidthplus} and \eqref{eq:linewidthminus}, which correspond to Eq. \eqref{eq:linewidth_resonantmodes} in the main text.\\

In the full device the far-detuned resonator $\Omega_1$ enters the input-output response through the same common waveguide, which creates an effective transmission-line-mediated interaction between $\Omega_1$ and $\Omega_{2,3}$. The complete model would be written as
\begin{equation}
    \hat{H}^{(3)}_{\text{eff}} = \left( \begin{array}{ccc}
    \Omega_1 - i \frac{\kappa_{\mathrm{cav},1}}{2} & G_{12} & G_{13}\\
    G_{12} & \Omega_2 - i\frac{\kappa_{\text{cav},2}}{2} & G_{\theta_3} \\
    G_{13} & G_{\theta_3} & \Omega_3 - i\frac{\kappa_{\text{cav},3}}{2} 
    \end{array} \right).
\end{equation}
where $G_{1j}$ is the total effective coupling between resonators $1$ and $j = 2,3$. To first order in the perturbation series, resonator 1 only produces self-energy corrections to the frequencies, linewidths, and effective coupling of the $(2,3)$ modes, of the order $\kappa_i \kappa_1 /(\Omega_1 - \Omega_i)$. Since $\Omega_1$ is far detuned and its coherent coupling to the active pair $(2,3)$ is suppressed, these corrections are negligible compared with the MHz-scale linewidths. We therefore extract $\tilde{\kappa}_{\pm}$ from the reduced two-resonator non-Hermitian Hamiltonian, while retaining the resonator 1 in the full input-output fit of $\vert S_{21}(\omega) \vert$.\\

\subsection{Dark state generation \label{sec:app_darkstate_twores}}

\begin{figure}
    \centering
    \includegraphics{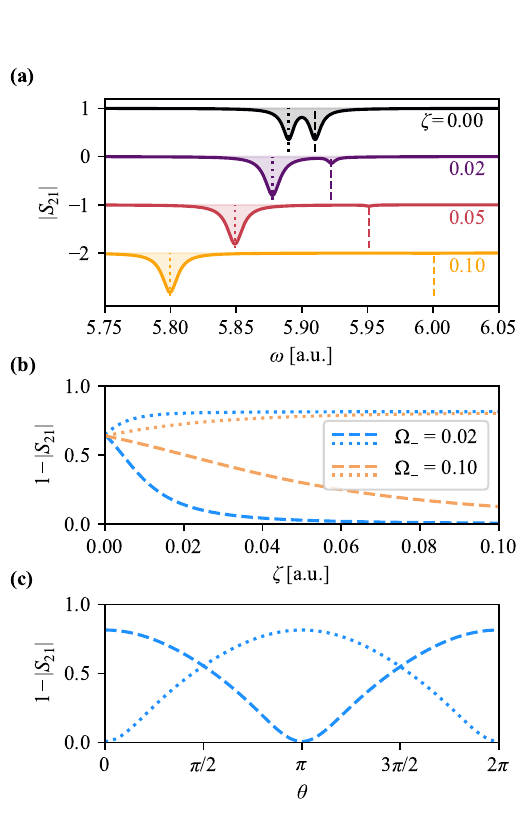}
    \caption{\textbf{Transmission coefficient $\vert S_{21} \vert$ as a function of different parameter choices, given fixed $\kappa_2 = \kappa_3 \in \mathbb{R}$ $(\phi_2 = \phi_3 = 0)$.} All parameters are expressed in the same arbitrary units of energy. 
     \textbf{(a)} $\vert S_{21} \vert$ as a function of the input frequency $\omega$, for a fixed resonator detuning $\Omega_- = 0.02$ and propagation phase $\theta_3 = \pi$. Vertical lines indicate the position of the peaks as predicted by Eq. \eqref{eq:twores_eig} (dashed for $\tilde{\Omega}_+$ and dotted for $\tilde{\Omega}_-$). An increasing value of $\zeta$ suppresses the spectral weight of one the photonic modes, rendering $\ket{\tilde{\psi}_+}$ dark to measurement. A vertical offset is added for clarity. \textbf{(b)} Spectral weight $1 - \vert S_{21}\vert$ , evaluated at the position of the peaks (dashed and dotted for the $+$ and $-$ mode, respectively) as a function of the resonator-resonator coupling $\zeta$, for two different resonator detunings (blue and orange) and fixed $\theta_3 = \pi$. For a given $\zeta$, the peak suppression ($1 - \vert S_{21}\vert \approx 0 $) is stronger for smaller detunings. \textbf{(c)} Spectral weight $1 - \vert S_{21}\vert$ , evaluated at the position of the peaks, for fixed detuning $\Omega_- = 0.02$ and $\zeta = 0.1$, as a function of the propagation phase $\theta_3$. Sweeping $\theta_3$ makes either $\ket{\tilde{\psi}_-}$ ($\theta_3 = 0, 2\pi$) or $\ket{\tilde{\psi}_+}$  ($\theta_3 = \pi$) dark to measurement.}
    \label{fig:two_res_darkstate}
\end{figure}

For the two coupled resonators, the jump operator to the right-moving channel can be written as 
\begin{equation}
    \hat{L}_+ = \sqrt{\kappa_3}\, \hat{a}_3 + e^{i\theta_3}\sqrt{\kappa_2}\,\hat{a}_2.
\end{equation}
By obtaining $\hat{L}_+\ket{\tilde{\psi}_+} = 0$, we can find the condition for $\ket{\tilde{\psi}_+}$ to be a dark state of the system:
\begin{equation}
    \sqrt{\kappa_3}\, \sin\left(\frac{\varphi}{2}\right) + e^{i\theta_3} \sqrt{\kappa_2} \cos\left( \frac{\varphi}{2}\right) = 0.
    \label{eq:darkstate_tworesonators_trans}
\end{equation}
This result allows us to analyze how different parameter choices affect the emergence of a dark state:
\begin{itemize}
    
    \item If both \(\{\kappa_2,\kappa_3 \} \in \mathbb{R}\), Eq.~\eqref{eq:darkstate_tworesonators_trans} becomes
    \begin{equation}
    \tan(\varphi/2) = -e^{i\theta_3}\sqrt{\kappa_2/\kappa_3},
    \label{eq:darkstate_cond_realkappa}
    \end{equation}
    which can only be satisfied for \(\theta_3 = \pm (2n-1)\pi\), with \(n \in \mathds{Z}_+\). For a narrow frequency window around a relevant input frequency \(\omega_0\), the dispersion relation of the waveguide can be approximated as \(k(\omega_0) = 2\pi/\lambda_0\). This implies that the separation between the resonators must satisfy
    \begin{equation}
    \Delta x = \frac{\vert \theta_3 \vert}{k(\omega_0)} = (2n-1)\frac{\lambda_0}{2}.
    \end{equation}
    \textit{We therefore find that exact darkness of the eigenstate \(\ket{\tilde{\psi}_+}\) requires the coupling points to be separated by odd multiples of \(\lambda_0/2\), together with compatible relative coupling strengths such that \(\tan(\varphi/2) = \sqrt{ \kappa_2 / \kappa_3 }\).} In the particular case \(\kappa_2 = \kappa_3\), the dark state emerges only for vanishing frequency detuning between resonators. This is similar to the conclusion of Refs.  \cite{Propp22, PhysRevA.79.042302}.
    
    \item Finally, if the coupling rates to the transmission line are allowed to carry phases, \(\kappa_j = |\kappa_j| e^{-i\phi_j}\), the dark-state condition on $\theta_3$ becomes
    \begin{equation}
    \theta_3 + (\phi_3 - \phi_2)/2 = \pm (2n-1)\pi.
    \end{equation}
\end{itemize}
with $n \in \mathds{Z_+}$. Figure~\ref{fig:two_res_darkstate} shows the transmission amplitude as a function of the relevant system parameters, illustrating how their interplay governs the emergence or disappearance of the dark state. For all panels, $\kappa_2$ and $\kappa_3$ are fixed to be equal and real. In Fig.~\ref{fig:two_res_darkstate}(a), for a fixed resonator detuning $\Omega_- = 0.02$ and propagation phase $\theta_3 = \pi$, one of the transmission peaks gradually loses spectral weight as the inter-resonator coupling is increased. The vertical lines (dotted and dashed) indicate the analytical prediction for the position of the peaks, as given by Eq.~\eqref{eq:twores_eig}. For $\zeta = 0.1$, the right peak (dashed, vertical lines) becomes invisible in transmission, signaling the dark state character of the corresponding photonic mode, $\ket{\tilde{\psi}_+}$.\\
Panel~\ref{fig:two_res_darkstate}(b) shows that increasing the resonator detuning to \(\Omega_- = 0.1\) requires a larger inter-resonator coupling strength to achieve complete suppression of the mode in transmission. The corresponding spectral weight is represented as \(1-|S_{21}|\), which vanishes when the peak is fully suppressed.\\  
Panel~\ref{fig:two_res_darkstate}(c) further illustrates how varying the propagation phase $\theta_3$ drives the system away from the condition for peak suppression, given $\Omega_- = 0.02$ and $\zeta = 0.1$. Analogous conditions to Eq. \eqref{eq:darkstate_cond_realkappa} can also be derived under which $\ket{\tilde{\psi}_-}$ becomes the dark state of the system, as is evident from the cases $\theta_3=0$ and $\theta_3 = 2\pi$ in panel \ref{fig:two_res_darkstate}(c). The right peak $\ket{\tilde{\psi}_+}$ is suppressed for $\theta_3 = \pi$, as expected.

The left-propagating channel constitutes an additional decay pathway, with associated jump operator $\hat{L}_- = \sqrt{\kappa_2} \, \hat{a}_2 + e^{i\theta_3 }\sqrt{\kappa_3} \, \hat{a}_3$, which imposes the following constraint on the system parameters for $\ket{\tilde{\psi}_+}$ to become a dark state, 
\begin{equation}
    \sqrt{\kappa_3}\, \sin\left(\frac{\varphi}{2}\right) + e^{-i\theta_3} \sqrt{\kappa_2} \cos\left( \frac{\varphi}{2}\right) = 0.
    \label{eq:twores_condition}
\end{equation}
For $\kappa_2,\kappa_3 \in \mathbb{R}$, this equation is again solved by $\theta_3 = \pm (2n-1)\pi$. In that case, the state is dark with respect to both output channels, i.e. it is completely invisible to the feedline because it does not radiate in either direction, and hence cannot be detected in either the transmitted or the reflected signal. For complex $\kappa_2,\kappa_3$, the condition for darkness to reflection reads $-\theta_3 + (\phi_3 - \phi_2)/2 = \pm (2n-1)\pi$. Requiring simultaneous darkness in transmission and reflection amounts to solving both $\hat{L}_\pm \ket{\tilde{\psi}_+}=0$, whose solution is shown in the main text. 

\section{Dark state generation in a two-coupled-resonator system with a spin ensemble}
Let us now consider the Hamiltonian describing the presence of a spin mode directly coupled to one of the resonators, 
\begin{equation}
    \left( \hat{\mathcal{H}}_{\text{spin}} + \hat{\mathcal{H}}_{\text{spin-cav}}  + \hat{\mathcal{H}}^{2,3}_{\text{cav-cav}} \right) /  \hbar = \left( 
    \begin{array}{ccc}
    \Omega_2 & \zeta & 0 \\
    \zeta & \Omega_3 & g \\
    0 & g & \Delta
    \end{array}
    \right).
    \label{eq:Hamiltonian_3LS}
\end{equation}
As a first step, we rewrite Eq.~\eqref{eq:Hamiltonian_3LS} in terms of the eigenstates of the coupled two-resonator subsystem, namely $\ket{\psi_+}$ and $\ket{\psi_-}$ shown in Eqs. \eqref{eq:eigenstates_plus} and \eqref{eq:eigenstates_minus}. In the basis $\{\ket{\downarrow, \psi_+},\ket{\downarrow, \psi_-}, \ket{\uparrow,0,0}\}$, the Hamiltonian takes the form
\begin{equation}
    \hat{\mathcal{H}}_{\text{s-r-r}} = \left( 
        \begin{array}{ccc}
              \tilde{\Omega}_+ - \Delta & 0 & g\sin\left(\frac{\varphi}{2}\right) \\
            0 &  \tilde{\Omega}_- -  \Delta  &  g\cos\left(\frac{\varphi}{2}\right) \\
            g\sin\left(\frac{\varphi}{2}\right)  &        g\cos\left(\frac{\varphi}{2}\right)  & 0
        \end{array}
    \right).
\end{equation}
In this representation, the light-matter interaction couples the spin excitation to both hybridized resonator modes. For generic values of $\zeta,\Omega_-$, neither $g\cos\left(\frac{\varphi}{2}\right)$ nor $g\sin\left(\frac{\varphi}{2}\right)$ can be neglected. As a consequence, the spin mediates an indirect coupling between the two modes $\ket{\tilde{\psi}_+}$ and $\ket{\tilde{\psi}_-}$. Therefore, even if the two-resonator subsystem is tuned to satisfy the dark-state condition, the inclusion of the spin generally spoils the perfect darkness of that mode.\\
Let us analyze some limiting case. First, for $\varphi \approx 0$ (namely, $\Omega_{-} \gg \zeta$) we can take $\cos(\varphi/2) \approx 1$ and $\sin(\varphi/2) \approx 0$, which simplifies the Hamiltonian to be
\begin{equation}
    \hat{\mathcal{H}}_{\text{s-r-r}}(\varphi\approx0 ) = \left( 
        \begin{array}{ccc}
            \tilde{\Omega}_+ -\Delta & 0 & 0 \\
            0 & \tilde{\Omega}_- - \Delta & g \\
            0  &       g           & 0
        \end{array}
    \right).
\end{equation}
As a result, the state $\ket{\downarrow,\tilde{\psi}_+}$ becomes effectively decoupled from the rest of the Hilbert space. Therefore, if $\ket{\tilde{\psi}_+}$ is tuned to satisfy the dark-state condition, the coupling to the spin ensemble does not spoil its perfect darkness. Of course, the condition \(\varphi \approx 0\) must also be compatible with Eq.~\eqref{eq:darkstate_cond_realkappa}. This, in turn, requires an imbalance in the coupling of the two resonators to the transmission line, namely \(\kappa_3 \gg \kappa_2\).\\

Let us now address the question of whether or not a hybridized eigenstate of the three-party system can turn into a dark state. Let us write a general eigenstates in the form
\begin{equation}
    \ket{\psi} = \alpha \ket{\downarrow, \tilde{\psi}_+} + \beta \ket{\downarrow, \tilde{\psi}_-} + \eta \ket{\uparrow,0,0},
\end{equation}
where $\alpha,\beta,\eta$ are the amplitudes to be determined. By operating 
\begin{align}
    \hat{L}_\pm \ket{\psi}  = & \sqrt{\kappa_3}\left[ \alpha \, \sin\left( \frac{\varphi}{2} \right)  +\beta \cos\left( \frac{\varphi}{2}\right) \right]\ket{\downarrow,0,0} \nonumber \\
    & +e^{\pm i\theta} \sqrt{\kappa_2}\left[ \alpha \cos\left( \frac{\varphi}{2} \right) - \beta \sin\left( \frac{\varphi}{2}\right)\right]\ket{\downarrow,0,0},
\end{align}
the condition for the dark state generation turns to be
\begin{align}
    \sqrt{\kappa_3}& \left[\alpha \sin\left( \frac{\varphi}{2}\right) + \beta \cos\left(\frac{\varphi}{2} \right)\right] \nonumber \\
    & +e^{\pm i\theta}\sqrt{\kappa_2}\left[ \alpha \cos\left(\frac{\varphi}{2} \right) - \beta \sin\left( \frac{\varphi}{2}\right) \right] = 0.
\end{align}
\begin{figure}[!b]
    \centering
    \includegraphics{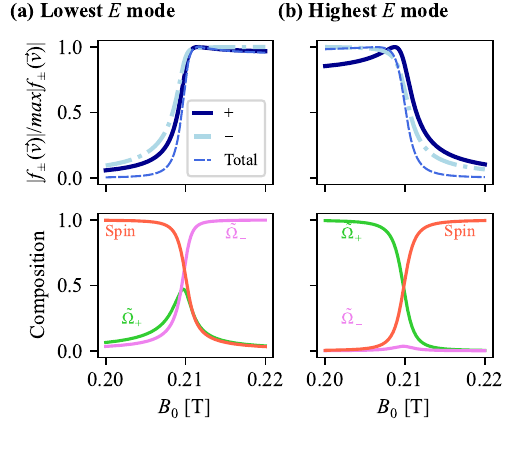}
    \caption{\textbf{Dark state condition for the lowest- and highest-energy mode.}  Dark state condition for the \textbf{(a)} lowest-energy eigenstate,  \textbf{(b)} highest-energy eigenstate, of the Hamiltonian in Eq. \eqref{eq:hamiltonian_spin_hybrized}, evaluated through $f_{\pm}(\vec{v}) / \max(f_{\pm}(\vec{v}))$, using the parameters $\vec{v} = (\kappa_2,\kappa_3,\alpha,\varphi, \theta)$ obtained from the fitting in Table \ref{tab:spin_hamiltonian}, as shown in the upper panels. The bottom panels show the eigenstate composition as a function of $B_0$. }
    \label{fig:composition_other_eigenstates}
\end{figure}

\section{Dark states in the three-mode system
\label{app:composition_other_eigenstates}}
In this section, we evaluate the dark-state condition for the lowest- and highest-energy modes using the experimental parameter values, as shown in Fig. \ref{fig:composition_other_eigenstates}. We find that neither mode supports the formation of a dark state. Their coupling to the transmission line is reduced only when they become predominantly spin-like and therefore have negligible weight in the photonic subspaces, as illustrated in the lower panels. Conversely, when the modes are predominantly photonic, their coupling to the transmission line is enhanced. \\

\section{Phase estimation due to the propagation delay between resonators \label{sec:app_phase_estimation}}
Within a microwave circuit, propagating modes exhibit a finite propagation time and accumulate an additional phase. 
This phase factor is essentially given by the distance between the resonators and the microwave frequency $\nu$ of the propagating signal, which we assume to be the resonance frequency of the corresponding resonator. 
In particular, adjacent resonators are distanced by $\Delta x_{j} = \vert x_j-x_{j-1}\vert = \SI{1500}{\micro\meter}$ within our superconducting circuit. We can determine the group velocity as
\begin{align}
    v_{\text{g}} = \frac{c}{\sqrt{\epsilon_{\text{eff}}/2}} \text{,}
\end{align}
with the speed of light $c$ and the effective permittivity of silicon $\epsilon_{\text{eff}}=11.7$. 
Consequently, we estimate $\theta_{2} = \theta_3 \approx  0.45$~rad, with a total accumulated phase of $\theta_3 + \theta_2 \approx 0.9$~rad over the length of the circuit.
While the estimate for $\theta_{2}$ is in good agreement for the fitted value presented in Tab.\,\ref{tab:spin_hamiltonian}, the estimated $\theta_{3}$ deviates from the fitted value. Here, we want to emphasize that our theoretical model assumes a point-like coupling of the microwave resonator to the transmission line waveguide. In our experimental circuit, however, the resonators have a finite width that is comparable to the distance between the resonators. Consequently, the capacitive coupling of the resonator to the waveguide is not point-like and the assumed distance $\Delta x_{j}$ varies from the assumed $\SI{1500}{\micro\meter}$.

%

\end{document}